\documentclass[12pt,tightenlines,superscriptaddress,showpacs,showkeys,amsmath,amssymb]{revtex4}

\usepackage{amsfonts}	
\usepackage{mathrsfs}	
\usepackage{graphicx}

\newcommand{\pa}[1]{\ensuremath{\left(#1\right)}}
\newcommand{\e}[1]{\ensuremath{{}_{\text{#1}}}}
\newcommand{\f}[2]{{\ensuremath{\mathchoice%
	{\dfrac{#1}{#2}}
        {\dfrac{#1}{#2}}
	{\frac{#1}{#2}}
	{\frac{#1}{#2}}
	}}}
\newcommand{\tf}[2]{\ensuremath{#1/#2}}
\newcommand{\U}[1]{\ensuremath{\mathrm{~#1}}}
\newcommand{\crochets}[1]{\ensuremath{\left[#1\right]}}
\newcommand{\pac}[1]{\ensuremath{\crochets{#1}}}		
\newcommand{\degre}{\ensuremath{{}^{\text{\footnotesize\ensuremath{\circ}}}}}
\newcommand{\degres}{\degre}

\begin{document}

\author{M. N. Bussac}
\affiliation{%
Laboratoire d'Opto\'electronique des Mat\'eriaux Mol\'eculaires\\
\'Ecole Polytechnique F\'ed\'erale de Lausanne, 1015 Lausanne,
Switzerland} 
\affiliation{Centre de Physique Th\'eorique\\
\'Ecole polytechnique, 91128 Palaiseau Cedex, France}

\author{J.-D. Picon}
\affiliation{%
Laboratoire d'Opto\'electronique des Mat\'eriaux Mol\'eculaires\\
\'Ecole Polytechnique F\'ed\'erale de Lausanne, 1015 Lausanne,
Switzerland} 

\author{L. Zuppiroli}
\email{libero.zuppiroli@epfl.ch}
\homepage[url: ]{lomm.epfl.ch}
\affiliation{%
Laboratoire d'Opto\'electronique des Mat\'eriaux Mol\'eculaires\\
\'Ecole Polytechnique F\'ed\'erale de Lausanne, 1015 Lausanne,
Switzerland}

\title{The Impact of Molecular Polarization on the Electronic
Properties of Molecular Semiconductors }

\date{\today}


\begin{abstract}
  In a molecular semiconductor, the carrier is dressed with a
  polarization cloud that we treat as a quantum field of Frenkel
  excitons coupled to it. The consequences of the existence of
  this electronic polaron on the dynamics of an extra charge in a
  material like pentacene can thus be evaluated.
\end{abstract}


\maketitle

Molecular organic semiconductors in general and oligoacenes in
  particular are seriously being considered today as materials for
  application in optoelectronic devices. Pentacene, for example, has
  already been used with success in organic field effect transistors
  \cite{Alenf,Nelson,Gundlach}. Critical to their eventual use as
  electronic materials, therefore, is a proper understanding of the
  charge transport mechanisms in these molecular solids. Currently, the
  paradigm in force describing the electronic properties of oligoacenes
  views carrier transport in terms of lattice polaron theories applied
  to narrow band crystals
  \cite{Warta,Kenkre,Giuggoli,Coropceanu,Cheng}. However, we believe,
  as other   authors \cite{Hill,Tsiper} have also pointed out, that
  this approach omits an important effect specific to organic
  materials composed of large polarizable molecules, i.e., adding a
  carrier onto a molecule in such a crystal, creates a polarization
  cloud around the carrier, also-called electronic polaron or Coulomb
  polaron \cite{Toyozowa}, which accompanies the itinerant charge as
  it moves about in the solid. This effect has actually been confirmed
  by photoemission experiments \cite{Pope,Song} wherein the spectra of
  individual molecules in the gas phase are compared to the
  corresponding spectra in the solid phase \cite{Pope} or within a
  molecular cluster \cite{Song}.   
  
  In these crystals, the time scale for establishing molecular
  polarization is related to the energetic distance $\varepsilon$
  between the ground state and the first excited state of the neutral
  molecule, generally on the order one electronvolt. The characteristic
  time, given by $\tau_\varepsilon = \tf{h}{\varepsilon}$ (where $h$ is
  the Planck constant), is thus on the order of $2.10^{-15}\U{s}$. This
  is a very fast process, about ten times faster than the
  characteristic time for carrier motion from molecule to molecule
  $\tau_J = \tf{h}{J}$, where $J\sim 0,1\U{eV}$ is the bare transfer
  integral between two adjacent molecular sites along the easiest
  direction of propagation in the crystal \cite{Cheng}. On this basis,
  one would thus expect the Coulomb polaron to readily follow the
  extra charge in its motion through the crystal without any
  significant scattering or friction. The general feeling among
  physicists therefore is that polarization effects are minor and that
  electron-phonon processes probably explain the bulk of transport
  phenomena in these materials.    

  We will show, however, that the electronic polaron does actually play
  an important role in the dynamics of an extra carrier, in particular,
  by (\textit{i}) significantly renormalizing the transfer integrals in
  the perfect crystal, (\textit{ii}) enhancing the effect of any type of
  disorder including thermal disorder, and (\textit{iii}) creating
  correlations between diagonal (on-site) disorder and non-diagonal
  (transfer integral) disorder. In contrast to early calculations
  (reviewed by Silinsh and $\check{\mathrm{C}}$ap\'ek \cite{Silinsh}),
  as well as more recent ones \cite{Tsiper}, which use a
  self-consistent approach in the limit of vanishing molecular overlap
  to compute the polarization energies, we wiIl present here a
  theoretical approach which accounts for the finite bandwidth effects
  by treating the entire solid quantum-mechanically. This is essential
  for studying the transfer of the itinerant carrier through the
  solid. The treatment of polarization here is applicable to all types
  of organic solids composed of large conjugated molecules, disordered
  or not and provides good insight into the possible roles of
  polarization in charge transport.   

  To demonstrate our points, we start from a model hamiltonian, inspired
  by the pioneering work of $\check{\mathrm{C}}$ap\'ek on the
  polarization of a molecular trimer \cite{Capek}. We will consider
  both the tight binding charge propagation from site $n$ to the
  neighboring sites $n + h$ and the consequent polarization of the
  molecular sites $\ell$ coupled electrostatically to the extra
  charge. Thus, the annihilation and creation operators $\hat{a}$ and
  $\hat{a}^+$ describe the narrow band $J$ excitations while the
  operators $\hat{b}$ and $\hat{b}^+$ take into account the internal
  electronic degrees of freedom in the neutral molecules responsible
  for polarization.  
    
  When the molecule is anisotropic, like pentacene, the polarization
  excitations are direction dependent. It is convenient then to describe
  each molecule $\ell$ located at position $\mathbf{r}_\ell$ in the
  frame of the principal axes of its polarizability tensor
  $\Bar{\Bar{\alpha}}_\ell$. In this orthonormal frame where the unit
  vectors can be called
  $(\mathbf{e}_\ell^1,\mathbf{e}_\ell^2,\mathbf{e}_\ell^3)$,  the
  dipolar transition tensor $\Bar{\Bar{\mu}}_\ell$ is also diagonal and
  can be represented by the three components ${\mu_\ell}_i$   
  
  \begin{equation}
    \mu_{\ell_i}=e\,\langle\phi_H|\mathbf{r}_{\ell_i}|\phi_L^i\rangle
    \label{mu}
  \end{equation}

  where $\phi_H$ is the non-degenerate HOMO state and $\phi_L^i$
  the particular LUMO state excited by a dipolar transition in the
  direction $i$. In the spirit of a tight binding model, the $\phi_L^i$
  are three different linear combinations of the $\pi$ orbitals selected
  when the particular direction $i$ of the dipolar transition is
  considered. These states have essentially the same energy
  $\varepsilon$ with respect to the HOMO level \cite{Glaeser}.  In this
  approach, the polarization of the molecules $\ell$  results from a
  small depopulation of the HOMO in favor of the LUMO. This excitation
  process, internal to the molecules is represented by the operator
  pairs $\hat{b}_{\ell_i}$ and $\hat{b}_{\ell_i}^+$.

  The quantum operator representing the dipolar moment on each site
  $\ell$ is the vector   
  \begin{equation}
    \mathbf{\hat{d}}_\ell=\sum_{i=1}^{3}\hat{d}_{\ell_i}
    \,\mathbf{e}_\ell^i \qquad \text{with} \qquad \hat{d}_{\ell_i}=\mu_{\ell_i}
    \,({\hat{b}_\ell+\hat{b}_\ell^+})_i 
  \end{equation}

  The diagonal polarization tensor $\Bar{\Bar{\alpha}}_\ell$ is then
  written as
  $(\Bar{\Bar{\alpha}}_\ell)_{ii}=\tf{2\,{{\mu_\ell}_i}^2}{\varepsilon}$.   
  \smallskip

  With this notation, the hamiltonian for the extra charge, in the
  presence of the polarization cloud can then be written in the
  dipolar approximation as:   
  \begin{eqnarray}
    \label{hamiltonien}
    \lefteqn{
      \hat{H}=
      \sum_{n}^{}\hat{a}_n^+\,\hat{a}_n\left\{ \sum_{\ell\neq
	n}^{}\,\sum_{i=1}^{3}\pac{\varepsilon\,(\hat{b}_\ell^+
	\,\hat{b}_\ell)_i-\f{q}{4\,\pi\,\varepsilon_0}
	\f{(\mu_\ell)_i\,
	  \pa{\mathbf{r}_\ell-\mathbf{r}_n}.\mathbf{e}_\ell^i\, 
	  (\hat{b}_\ell^++\hat{b}_\ell)_i}
	  {|\mathbf{r}_\ell-\mathbf{r}_n|^3}}\right.}\\ \nonumber
    &&   \left.+\f{1}{2}\sum_{\ell\neq m}^{}
    \sum_{m}^{}\sum_{i,j=1}^{3}\pa{W_{\!\ell,\,m}}_{\!i,\,j}\,
    \pa{\hat{b}_\ell^++\hat{b}_\ell}_{\!i} \,
    \pa{\hat{b}_m^++\hat{b}_m}_{\!j}\right\}
    -\sum_{n}^{}\sum_{h}^{} J_{n,n+h}\,\hat{a}_{n+h}^+\,\hat{a}_n 
  \end{eqnarray}

  where $J_{n, n+h}$ are the transfer integrals between sites $n$ and
  $n+h$ and 
  \begin{equation}
    \pa{W_{\ell,m}}_{i,j}=\f{\mu_{\ell_i}\,\mu_{m_i}}{4\,\pi\,\varepsilon_0 
      \,|\mathbf{r}_\ell-\mathbf{r}_m|^3}\pac{
       (\mathbf{e}_\ell^i.\mathbf{e}_m^j)-
      3\f{(\mathbf{r}_\ell-\mathbf{r}_m).
      \mathbf{e}_\ell^i   
	\,(\mathbf{r}_\ell-\mathbf{r}_m).\mathbf{e}_m^j}
      {|\mathbf{r}_\ell-\mathbf{r}_m|^2} 
    }
    \label{wdipoldipol}
  \end{equation}

  are the dipolar interactions between pairs of molecules in the
  structure, responsible for the Van der Waals contribution to the
  cohesive energy.   

  In fact this hamiltonian describes a simplified version of the
  motion   of a charge carrier interacting with a quantum field of
  Frenkel   excitons of energy $\varepsilon$ \cite{Toyozowa}.  

  In molecular crystals like acenes, the motion of the excess charge
  is slow compared to the relaxation time necessary for the
  polarization of the electronic orbitals of the molecules surrounding
  the charge carrier, i.e. $J\ll\varepsilon$. To approximate ground
  state of the hamiltonian \ref{hamiltonien}, we have thus used a
  variational method with a dressed carrier trial function. A
  perturbation theory would also give such solution when
  $\tau_\varepsilon\ll \tau_J$. The eigenstates of the hamiltonian are
  represented as a superposition of local states, taken to be the
  product of a local electronic state $|n\rangle$ and the
  «polarization» state $|\chi (n)\rangle$ of the surroundings,
  associated with the carrier's occupation of site $n$ :  

  \begin{equation}
    |\psi\rangle=\sum_{n}^{}u_n\,|n\rangle\otimes|\chi(n)\rangle
    \label{psi}
  \end{equation}

  The local polarization states which constitute $|\chi (n)\rangle$
  can be expressed in terms of the unitary translation operators
  $\hat{U}_\ell$ where $|0_\ell\rangle$ is the ground state of the
  $\ell$-th molecule and

  \begin{eqnarray}
    \label{chi}
    |\chi (n)\rangle & = & \otimes\prod_{\ell}^{}\hat{U}_\ell(n)\,
    |0_\ell\rangle\\  \nonumber 
    \hat{U}_\ell(n)  & = & \exp{\pa{
	\sum_{i=1}^{3}\pa{X_{\ell,i}^*(n)\,\hat{b}_{\ell_i}
	  -X_{\ell,i}(n)\,\hat{b}_{\ell_i}^+}  
    }}
  \end{eqnarray}

  The unitary operator $\hat{U}_\ell(n)$ represents a translation of
  the molecular state due to the electric field of the charge placed
  on site $n$:

  \begin{eqnarray}
    \hat{U}_\ell^{-1}(n)\,\hat{b}_{\ell_i}\,\hat{U}_\ell(n) &=&
    \hat{b}_{\ell_i}-X_{\ell,i}(n)\,\hat{I}_\ell\\ [3mm] \nonumber
    \hat{U}_\ell^{-1}(n)\,\hat{b}_{\ell_i}^+\,\hat{U}_\ell(n) &=&
    \hat{b}_{\ell_i}^+-X_{\ell,i}^*(n)\,\hat{I}_\ell
  \end{eqnarray}
  
  where $\hat{I}_\ell$ is the identity operator.
  
  The functions $X_{\ell,i}(n)$ are determined by minimizing the
  expectation value of the  Hamiltonian \ref{hamiltonien} with the
  variational wave function \ref{psi} and \ref{chi} above :  

  \begin{eqnarray}
    \lefteqn{
      \langle\psi|\hat{H}|\psi\rangle = \sum_{n}^{} |u_n|^2\left\{
      \sum_{\ell\neq n}^{}\sum_{i=1}^{3}\pa{ \varepsilon|X_{\ell,i}(n)|^2 +
	\f{q}{4\pi\varepsilon_0}\mu_{\ell_i}
	\f{(\mathbf{r}_\ell-\mathbf{r}_n).\mathbf{e}_\ell^i}
	  {|\mathbf{r}_\ell-\mathbf{r}_n|^3}
	  \pa{X_{\ell,i}(n)+X_{\ell,i}^*(n)}}\right.}\\ \nonumber 
    && \left. +\f{1}{2} \sum_{\ell\neq m}^{}\sum_{m}^{}\sum_{i,j=1}^{3}
    \pa{W_{\ell,m}}_{i,j}\pa{X_{\ell,i}(n)+X_{\ell,m}^*(n)}
    \pa{X_{m,j}(n)+X_{m,j}^*(n)}\right\}\\\nonumber
    && -\sum_{n}^{}\sum_{h}^{}J\,u_{n+h}^*\,u_n \exp{\pa{
	-\f{1}{2}\sum_{\ell}^{}\sum_{i=1}^{3}\pa{
	  |X_{\ell,i}(n)|^2+|X_{\ell,i}(n+h)|^2-2X_{\ell,i}(n)X_{\ell,i}^*(n+h)
    }}}
    \label{psihpsi}
  \end{eqnarray}
  
  where we have used the Weyl identity 

  \begin{equation}
    \langle0_\ell|\,\text{e}^{\pa{\delta\,\hat{b}_\ell^+-\delta^+\hat{b}_\ell}}
    \,\text{e}^{\pa{\gamma^+\hat{b}_\ell-\gamma\,\hat{b}_\ell^+}}\,
    |0_\ell\rangle=   
    \text{e}^{\pa{\delta^+\gamma-\tf{(|\delta|^2+|\gamma|^2)}{2}}}
  \end{equation}

  We now minimize the energy of equation \ref{psihpsi} with respect to
  the variational functions, the $X_{\ell,i}(n)$'s. Since we are
  concerned with the stability of a slow carrier, we  need only
  minimize the potential-like energy. Thus to the lowest order in
  $\tf{J}{\varepsilon}$, we get  

  \begin{equation}
    \varepsilon X_{\ell,i}(n)+\f{q}{4\pi\varepsilon_0}\mu_{\ell_i}
    \f{(\mathbf{r}_\ell-\mathbf{r}_n).\mathbf{e}_\ell^i}
      {|\mathbf{r}_\ell-\mathbf{r}_n|^3} +\sum_{m}^{}\sum_{j=1}^{3}
      \pa{W_{\ell,m}}_{i,j} \pa{X_{m,j}(n)+X_{m,j}^*(n)}=0
      \label{classique}
  \end{equation}
  
  Substituing \ref{classique} in \ref{psihpsi}, the local polarization
  energy becomes: 

  \begin{equation}
    E\e{p}(n)=\sum_{\ell\neq n}^{}\f{q}{4\pi\varepsilon_0}\sum_{i=1}^{3}
    \mu_{\ell_i} \f{(\mathbf{r}_\ell-\mathbf{r}_n).\mathbf{e}_\ell^i}
       {|\mathbf{r}_\ell-\mathbf{r}_n|^3}\, X_{\ell,i}(n)
       \label{energie}
  \end{equation}
  
  Within the subspace of states $|\psi\rangle$ representing the
  dressed-electrons defined by relation \ref{chi} , the total energy
  minimum \ref{psihpsi} can thus be written as:    

  \begin{equation}
    E=\sum_{n}^{}E\e{p}(n)\,|u_n|^2-\sum_{n,n+h}^{}
    \tilde{J}_{n,n+h}\,u_{n+h}^*\,u_n  
    \label{energietotale}
  \end{equation}

  Then, the hamiltonian \ref{hamiltonien} describes a dressed carrier
  (or a quasi-particle) at site $n$ having a potential energy $E\e{p}
  (n)$ and an effective transfer integral   

  \[
  \tilde{J}_{n,n+h}=J_{n,n+h}\,\text{e}^{-S_0(h)}
  \]

  with

  \begin{equation}
    S_0(h)=\f{1}{2}\sum_{\ell}^{}\sum_{i=1}^{3}
    \pa{X_{\ell,i}(n)-X_{\ell,i}^*(n+h)}^2\cong
    \f{\gamma}{4}\f{E\e{p}(0)+E\e{p}(h)}{\varepsilon}  
    \label{so}
  \end{equation}

  where $\gamma$ is a coefficient of order unit.  

  The effective quasi-particle  hamiltonian $\hat{H}$ with parameters
  renormalized by the Frenkel exciton field is just the tight-binding
  hamiltonian : 

  \begin{equation}
    \hat{H}=\sum_{n}^{}E\e{p}(n)\,|n\rangle\langle n|-\sum_{n,n+h}^{}
    \tilde{J}_{n,n+h}\,|n\rangle\langle n+h|
    \label{liaisonf}
  \end{equation}

  where $E\e{p}(n)$ and $\tilde{J}_{n,n+h}$ are defined by relations
  \ref{energie} and \ref{so}.  

  In the case of a perfect crystal, the polarization energy $E\e{p}$
  is uniform and shifts the ground state everywhere by about $1\U{eV}$
  as observed in acenes \cite{Pope,Song}. Correlatively the bare
  bandwidth is significantly narrowed (independently of the
  temperature). Values of $E\e{p}$ (from \cite{Pope}) and
  $\tf{\tilde{J}}{J}$ have been reported in table \ref{tableau}. We
  can conclude therefore that at low temperatures in a perfect
  molecular crystal, a band model with an appropriate renormalized
  bandwidth describes the extended ground state properly
  \cite{Silinsh} (claim (\textit{i}) above). 

  The situation changes greatly with disorder, however small, enters
  the system. The local changes of polarization  energy $E\e{p}$ due
  to disorder are greater than any other electronic parameter change,
  transfer integral, HOMO and LUMO positions, etc. In particular,
  thermal disorder arising from librations or low energy
  intermolecular phonons is interpreted as being static on the time
  scale $\tau_J$ and induces non negligible polarization energy
  variations.  

  Indeed, consider any kind of static (or thermal) disorder which can
  be described by a random variable in the lattice. The polarization
  energy $E\e{p}$ is in turn also random. We denote $\delta_n$ the
  fluctuation of $E\e{p}$ from site to site. Then $\tilde{J}_{n, n+h}$
  also becomes random through the fluctuations of
  $X_{\ell,i}(n)$. Thus the motion of the quasi-particle representing
  an extra-charge in the disordered crystal is controlled by an
  Anderson hamiltonian with correlated diagonal and non-diagonal
  disorder that we shall now study in more detail.   

  \begin{figure}[ht]
    \begin{center}
      \includegraphics[width=8cm]{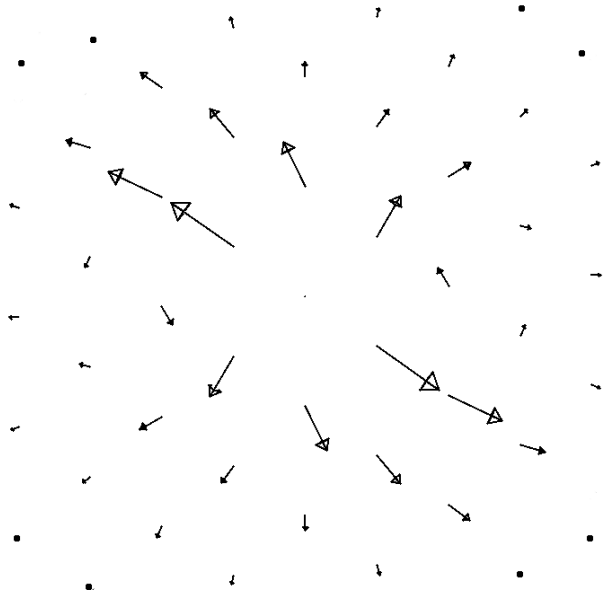}
    \end{center}
    \caption{Distribution of molecular dipoles induced by a charge
      situated at the origin within the $(a,b)$ plane. The arrows are
      projections of the dipole vectors on the plane.}
    \label{cartedip}
  \end{figure}
  
  For this purpose we have developed a numerical method for
  calculating $E\e{p}$ and $\tilde{J}$ in disordered situations. Still
  working within the dipolar approximation, we solve the linear
  equations in relation \ref{classique} numerically. This calculation
  is precise enough to determine the characteristics of the electronic
  polaron in naphtalene and anthracene. In tetracene and pentacene
  where the molecular polarizability is higher, corrections beyond the
  dipolar approximation are important. In this case, the linear
  calculation would give values of $X_{\ell,i}(n)$ exceeding
  $\tf{1}{2}$ on certain sites, which would correspond to  values of
  the dipole components exceeding the corresponding dipolar transition
  component $\mu_{\ell_i}$ of relation \ref{mu}. In order to limit
  this  artefact and to take into account the hyperpolarizability, the
  terms $\varepsilon\, |X_{\ell,i}(n)|^2$ in equation \ref{psihpsi}
  are replaced by
  $\f{\varepsilon}{2}\pa{1-\sqrt{1-4|X_{\ell,i}(n)|^2}}$.  The system
  of equations in \ref{classique} then becomes nonlinear and  must be
  solved by using an iterative numerical  procedure. Figure
  \ref{cartedip} represents the dipole distribution induced by a
  charge in a perfect crystal of pentacene.     

  \begin{table}[ht]
    \caption{Polarization energy $E\e{p}$ and renormalized transfer
      integral $\tf{\tilde{J}}{J}$ in the perfect crystals of
      naphtalene, anthracene, tetracene and pentacene for an angular
      disorder of $3\degre$.}     
    \begin{center}
      \begin{tabular}{|lc|cccc|}\cline{3-6}
	\multicolumn{2}{l|}{} &\multicolumn{1}{c}{Nph}
	&\multicolumn{1}{c}{Ac} &\multicolumn{1}{c}{Tc} &
	\multicolumn{1}{c|}{Pc}\\\hline 
	&&  21,5  & 33,9  & 48,2  & 91    \\
	Polarization tensor $\Bar{\bar{\alpha}}$&$\U{(\AA^3)}$ &  17,6  &
	29,2  & 34,7  & 38    \\ 
	&&  10,1  & 12,9  & 15,6  & 17,7  \\ \hline
	Polarization energy$^{[20]}$ $E\e{p}$&$\U{(eV)}$ &  -0,99 & -1,19
	& -1,39 & -1,55 \\ \hline 
	Bare tranfer integral$^{[21]}$ $J$&$\U{(meV)}$ &  41,5  & 47,9  &
	68,8  & 97.8 
	\\ \hline 
	Renormalized transfer $\tf{\tilde{J}}{J}$& &  0,86  & 0,83  &
	0,74  & 0,64  \\ \hline 
	Diagonal disorder $\delta$&$\U{(meV)}$ &  6,2   & 9,7   & 11,3  & 15,8
	\\ \hline 
	Correlation factor $\delta '$&$\U{(eV)}$ &  5,3   & 3,4   & 1,0   & 0,4
	\\ \hline 
      \end{tabular}
    \end{center}
    \label{tableau}
  \end{table}
  
  We have also introduced an orientational disorder which is known to
  be important in acenes. Pope and Swenberg give typical thermal
  disorder values for rotation angles of $\pm 3$ degrees at room
  temperature \cite{Silinsh2}. Other sources of rotational disorder
  with angles even larger than $3\degres$ are also likely in actual
  acene materials, for example, metastable polymorph domains, stacking
  faults, zones close to grain boundaries, dislocation cores
  etc. \cite{Brinkmann}. The simplest disorder distribution that is
  worth investigating is an angular gaussian distribution with mean
  square value $\Delta\theta$. By averaging over $4 000$ samples for
  each $\Delta\theta$, the mean square fluctuations   

  \[
  \delta =\sqrt{\langle {E\e{p}}^2-\langle E\e{p}\rangle^2\rangle}
  \]

  and $\delta '$ can be deduced as functions of $\Delta\theta$. Table
  \ref{tableau} summarizes the  results of our
  calculations. Polarization energy $E\e{p}$ and renormalized transfer
  integral $\tf{\tilde{J}}{J}$ in the perfect crystal are calculated
  from equations \ref{energie} et \ref{so} respectively. The diagonal
  energy disorder $\delta$ is calculated for an angular disorder of
  $3\degre$ while $\delta '$ represents the correlations between
  diagonal and nondiagonal disorder (see equation \ref{anderson}). The
  polarization tensors and the bare transfer integrals are from
  litterature.\footnote{The values for Nph, Ac and Tc are from
  \cite{Silinsh}; the values for Pc are from \cite{Soos}.} \footnote{The
  values are from \cite{Cheng}, along the easiest propagation
  direction.}  

  On the basis of this result, a good approximation for the related
  Anderson hamiltonian, would be    

  \begin{equation}
    \hat{H}=\sum_{n}^{}\delta_n|n\rangle\langle n|-\sum_{n,n+h}^{}
    \text{e}^{\tf{(\delta_n+\delta_{n+h})}{\delta'}}|n\rangle\langle n+h|
    \label{anderson}
  \end{equation}

  where the distribution of $\delta_n$ is characterized by the
  mean square value $\delta$ and the correlations by the value
  $\delta'$. In the two dimensional channel of an organic field effect
  transistor these disorder energies, reported in table \ref{tableau},
  are non-negligible fractions of the renormalized transfer integral
  $\tilde{J}$ .  
  
  It is important to note that from the point of view of the
  extra-charge traveling into the molecular semiconductor, any thermal
  or static disorder which modulates the bare tight binding parameters
  is amplified by molecular polarization. These «solvation» effects
  yield much larger disorder energies than the corresponding
  single-electron quantities. The amplification factor is of the order
  $\tf{E\e{p}}{J}\sim 10$ (claim (\textit{ii}) above). Moreover the electronic
  polaron effect introduces long-range correlations which are very
  efficient in terms of localization (claim (\textit{iii}) above). Work on the
  spectrum of hamiltonian \ref{anderson} is presently in progress to
  examine in detail the consequences of these correlations. At the
  present stage, we believe that the possibility to map the many-body
  hamiltonian \ref{hamiltonien} on the single electron hamiltonian
  \ref{liaisonf}, in the subspace of the class of pertinent wave
  functions, will be useful for further theoretical developments in the
  field.  
  
\acknowledgments{The authors acknowledge the help of the Swiss Federal
  Science Foundation under contract number 20-67929.02. We thank
  W. Leo for many useful comments.}

\bibliographystyle{unsrt}
\bibliography{biblio}

\begin{thebibliography}{10}

\bibitem{Alenf}
C.D. Dimitrakopoulos and R.L. Malenfant.
\newblock {\em Adv. Mat.}, 14:99, 2002.

\bibitem{Nelson}
S.F. Nelson, Y.-Y. Lin, D.J. Gundlach, and T.N. Jackson.
\newblock {\em Appl. Phys. Lett.}, 72:1854, 1998.

\bibitem{Gundlach}
D.J. Gundlach, J.A. Nichols, L.~Zhou, and T.N. Jackson.
\newblock {\em Appl. Phys. Lett.}, 80:2925, 2002.

\bibitem{Warta}
W.~Warta and N.~Karl.
\newblock {\em Phys. Rev. B}, 32:1172, 1977.

\bibitem{Kenkre}
V.M. Kenkre, J.D. Andersen, D.H. Dunlap, and C.B. Duke.
\newblock {\em Phys. Rev. Lett.}, 62:1165, 1989.

\bibitem{Giuggoli}
L.~Giuggoli, J.D. Andersen, and V.M. Kenkre.
\newblock {\em Phys. Rev. B}, 67:045110, 2003.

\bibitem{Coropceanu}
V.~Coropceanu, M.~Malagoli, D.A. da~Silva~Filho, N.E. Gruhn, T.G. Bill, and
  J.L. Bredas.
\newblock {\em Phys. Rev. Lett.}, 89:275503, 2002.

\bibitem{Cheng}
Y.C. Cheng, R.J. Silbey, D.A. da~Silva~Filho, J.P. Calbert, J.~Cornil, and J.L.
  Bredas.
\newblock {\em J. Chem. Phys.}, 118:3764, 2003.

\bibitem{Hill}
T.G. Hill, A.~Kahn, Z.G. Soos, R.A. Pascal, J.~Cornil, and J.L. Bredas.
\newblock {\em Chemical Physics Letters}, 327:181, 2000.

\bibitem{Tsiper}
E.V. Tsiper and Z.G. Soos.
\newblock {\em Phys. Rev. B}, 64:195124, 2001.

\bibitem{Toyozowa}
Y.~Toyozowa.
\newblock {\em Prog. Theor. Phys.}, 12:421, 1954.

\bibitem{Pope}
M.~Pope and C.E. Swenberg.
\newblock {\em Electronic Processes in Organic Crystals and Polymers}, page
  553.
\newblock Oxford University Press, New-York, 1999.

\bibitem{Song}
J.K. Song, S.Y. Han, I.~Chu, J.H. Kim, S.A. Kim, S.K.and~Lyapustina, S.~Xu,
  J.M. Nilles, and K.H. Bowen.
\newblock {\em J. Chem. Phys.}, 116:4477, 2002.

\bibitem{Silinsh}
E.A. Silinsh and V.~$\check{\mathrm{C}}$ap\'ek.
\newblock {\em Organic Molecular Crystals}, pages 101--125.
\newblock AIP Press, New-York, 1994.

\bibitem{Capek}
V.~$\check{\mathrm{C}}$ap\'ek.
\newblock {\em Czech J. Phys.}, B28:773, 1978.

\bibitem{Glaeser}
R.M. Glaeser and R.S. Berry.
\newblock {\em J. Chem. Phys.}, 44:3797, 1966.

\bibitem{Silinsh2}
E.A. Silinsh and V.~$\check{\mathrm{C}}$ap\'ek.
\newblock {\em Organic Molecular Crystals}, page 110.
\newblock AIP Press, New-York, 1994.

\bibitem{Brinkmann}
M.~Brinkmann, S.~Graff, C.~Straupe, J.C. Wittmann, C.~Chaumont, F.~N\"uesch,
  A.~Aziz, M.~Schaer, and L.~Zuppiroli.
\newblock {\em J. Chem. Phys. B.}, 107:10531--10539, 2003.

\bibitem{Soos}
E.V. Tsiper and Z.G. Soos.
\newblock {\em Phys. Rev. B}, 68:085301, 2003.

\end{thebibliography}

\end{document}